\documentclass[%
reprint,
unsortedaddress,
nofootinbib,
amsmath,amssymb,
aps,
prl,
]{revtex4-2}

\usepackage{graphicx}
\usepackage{amssymb,amsmath}
\usepackage[utf8]{inputenc}
\usepackage{multirow}
\usepackage{xcolor}
\usepackage{hyperref}
\hypersetup{colorlinks=true,linkcolor=blue,anchorcolor=blue,citecolor=magenta,filecolor=blue,urlcolor=blue,bookmarksnumbered=true}
\usepackage[flushleft]{threeparttable}
\usepackage{mathtools}

\usepackage{dsfont}
\usepackage{amsthm}

\usepackage{enumitem}
\setlist[itemize]{leftmargin=10mm}

\usepackage{multirow}

\usepackage{braket}

\usepackage{mathrsfs}

\usepackage{booktabs}

\begin{document}


\title{Ab initio many-fermion structure calculations on a quantum computer
} 

\author{Weijie Du\textsuperscript{1,2}}

\author{Yangguang Yang\textsuperscript{1}}

\author{Zixin Liu\textsuperscript{1}}
\email[Email: ]{zixinliu@gdlhz.ac.cn}

\author{Chao Yang\textsuperscript{3}}

\author{James P. Vary\textsuperscript{2}}

\affiliation{\textsuperscript{1}Institute of Modern Physics, Chinese Academy of Sciences, Lanzhou 730000, China}
\affiliation{\textsuperscript{2}Department of Physics and Astronomy, Iowa State University, Ames, Iowa 50010, USA}
\affiliation{\textsuperscript{3}Applied Mathematics and Computational Research Division, Lawrence Berkeley National Laboratory, Berkeley, California 54720, USA}

\date{\today}

\begin{abstract}

To overcome the limitations of existing algorithms for solving self-bound quantum many-body problems -- such as those encountered in nuclear and particle physics -- that access only a restricted subset of energy levels and provide limited structural information, we introduce and demonstrate a novel quantum-classical approach capable of resolving the complete bound-state spectrum. This method also provides the total angular momentum $J$ associated with each eigenstate.
Our approach is based on expressing the Hamiltonian in second-quantized form within a novel input model combined with a scan scheme, enabling broad applicability to configuration-interaction calculations across diverse fields. 
We apply this hybrid method to compute, for the first time, the bound-state spectrum together with corresponding $J$ values of ${^{20}O}$ using a realistic strong-interaction Hamiltonian.
Our approach applies to hadron spectra and $J$ values  solved in the relativistic Basis Light-Front Quantization approach.

\end{abstract}

\maketitle

{\it Introduction.--}
Ab initio, or first principles, Hamiltonian many-body calculations are widely employed in diverse fields such as quantum chemistry \cite{cook2012handbook,bauer2020quantum}, nuclear physics \cite{RevModPhys.77.427,RevModPhys.87.1067,Hjorth-Jensen:2017gss,Barrett:2013nh}, condensed matter physics \cite{fradkin2013field,michael2005computational}, and quantum field theories \cite{peskin2018introduction,BRODSKY1998299}. Classical computing faces the daunting challenge of exponentially increasing resources needed to obtain practical results, for example, to test the underlying interactions or search for novel phenomena. Recent advances in quantum computing have unlocked  exciting opportunities for performing efficient first-principles many-body calculations \cite{feynman1982simulating,lloyd1996universal,Cao_2019,McArdle_2020,RevModPhys.90.015002} but practical algorithms for the full spectra using real-space interactions have yet to emerge. We introduce a general-purpose framework and adopt the self-bound nucleus as an illustrative example to demonstrate how it provides access to the full spectra including determination of the $J$ values with strongly-interacting real-space interactions. 

Several quantum algorithms have been adopted to quantum many-body structure calculations, such as the variational quantum eigensolver \cite{Cao_2019,McArdle_2020,PhysRevLett.120.210501,Tilly_2022,Romero_2022,Sarma_2023,PhysRevC.106.034325,P_rez_Obiol_2023}, imaginary time quantum eigensolver \cite{motta2020determining,Yeter-Aydeniz:2019htv}, and quantum subspace expansion methods \cite{Cortes_2022,Epperly_2022,Kirby_2023,motta2023subspace,Stair_2020,Du:2024ixj,Du:2024zvr}. 
However, these algorithms primarily target the energies of a limited set of states, mostly the ground state, and struggle to access other structural information essential for understanding a system's response to external probes \cite{PhysRevLett.68.3682,Carlson:2014vla,PhysRevC.91.062501,PhysRevLett.117.082501,PhysRevC.90.064619,PhysRevLett.127.072501,PhysRevC.109.025502,acharya202416oelectroweakresponsefunctions,sobczyk2024spinresponseneutronmatter,EFROS1994130,Roggero:2018hrn,Yin:2024xsg,kurkcuoglu2025inferenceresponsefunctionshelp}, nuclear reactions \cite{Stetcu:2025mjw,weiss2025solvingreactiondynamicsquantum,Efros_2007,PhysRevC.99.034620,goldberger2004collision,hjorth2017advanced,PhysRevC.87.034326,Navratil_2009}, and scattering dynamics \cite{PhysRevC.82.034003,Du:2018tce,Du:2020glq,Sharma_2024,Wang:2024scd,Sharaf:2024vcm,Zhang:2019cai,Zhang:2020rhz,Zhang:2024ril,Zhang:2024gac}.

Meanwhile, systematically inputting many-fermion Hamiltonians into quantum computers is challenging. Especially, for the Hamiltonians including many-body interactions (e.g., in nuclear and particle physics), the Jordan-Wigner (JW) \cite{JordanWigner1928,Nielsen2005TheFC}  and Bravyi-Kitaev (BK) \cite{Bravyi_2002,Seeley_2012} schemes become inefficient, as compiling such terms in the Hamiltonian incurs significant overhead and produces extensive Pauli strings.
Oracle-based approaches, such as the linear combination of unitaries (LCU) \cite{Childs2012HamiltonianSU,PhysRevLett.114.090502}, also face growing challenges in designing practical oracles to extract numerous Hamiltonian matrix elements and to uncompute ancilla qubits.

We introduce a new quantum-classical method for {\it ab inito} structure calculations of many-fermion systems.
This approach incorporates a unique two-fold scan scheme that systematically resolves the eigenenergies and corresponding $J$ values of the Hamiltonian eigenstates.
For the first time, we compute the full spectrum, along with the corresponding $J$ values, of ${^{20}O}$ using quantum algorithms based on a realistic strong-force interaction \cite{Shin:2024zpe} developed from fundamental theories \cite{Epelbaum:2008ga,Epelbaum:2014efa,Epelbaum:2014sza}. The same method applies directly to relativistic Hamilonians derived from quantum field theory and expressed in the Basis Light-Front Quantization (BLFQ) framework \cite{Vary:2009gt,Qian:2020utg,Kreshchuk:2020dla,Kreshchuk:2020aiq,Du:2019ips,Du:2023bpw}.

As a crucial ingredient of this method, we propose a new Hamiltonian input scheme based on a novel circuit representation of the fermion operators that enforces constraints of valid many-body configurations in configuration interaction (CI) calculations.
This input scheme eliminates the overhead associated with compiling fermionic operators into Pauli strings that is required by the JW and BK transformations.
It is also oracle-free, and there is no need for any uncomputations. Additionally, this input scheme preserves Hamiltonian symmetries, a valuable feature to efficiently pruning Hilbert spaces for tailored eigenvalue calculations.

In the following, we begin with our general discussion of the formalism, Hamiltonian input scheme, and hybrid method. Then, we present the structure calculation of ${^{20}}O$ using our method. 
The proposed method is easily seen to be applicable to CI calculations across various fields.

{\it Formalism.--}
The Green's function formalism is a forefront research topic in classical computing \cite{schirmer2018many,stefanucci2025nonequilibrium,fetter2003quantum} as well as quantum computing \cite{Kokcu:2023vwg,Jensen_2023,PhysRevA.104.032422,keen2021,PhysRevResearch.4.043038,PhysRevResearch.4.023219,PhysRevResearch.2.033281}.
We can calculate the retarded Green's function $G(E) = 1/(E-H)$ (referred to as the resolvent), where it is understood that $H$ includes an infinitesimal imaginary part $H \rightarrow H - i \epsilon $ with $\epsilon > 0$, and that $||H||_2 \leq 1$ (one rescales $H$ otherwise). The resolvent can be expressed as \cite{PhysRevC.79.044308}:
\begin{equation}
	G(E) = - i \int ^{\infty} _{-\infty}  \exp[iEt]  \exp [-iHt] \ dt .
\end{equation}
We approximate $\exp [-iHt]$ with the Chebyshev polynomial (CP) expansion \cite{10.1063/1.448136,RevModPhys.78.275}
\begin{equation}
	\exp [-i H t] = \sum _{n=0}^{\infty} (-i)^n (2- \delta _{n,0}) J_n(t) T_n (H) ,
	\label{eq:expansion_time_evolution}
\end{equation}
where $J_n(\cdot)$ is the Bessel function of the first kind and $T_n(\cdot)$ denotes the CP of the first kind \cite{kenken2013mathematical}. 

We evaluate the integral using a discrete Fourier transform \cite{PhysRevC.79.044308,kenken2013mathematical}. With $N$ grid points, the time is given by $t = \pi \tau$ where $\tau = 0, 1, \cdots, N-1 $. The discrete energy points are $E_p = 2p/N$, where $p = -N/2, \cdots , N/2$, with $N$ being even.  
We have 
\begin{multline}
		\langle \psi _{\rm out } |  G(E_p) |  \psi _{\rm in } \rangle  = -i \pi \Bigg[ \sum _{\tau =0} ^{N-1} e^{2\pi i p \tau /N} \\
		 \times \sum _{n =0} ^{n_{\rm max} (\tau ) } (-i)^n (2-\delta_{n,0})  J_n(\pi \tau) \langle \psi _{\rm out} | T_n (H) | \psi _{\rm in} \rangle  \Bigg] .
		 \label{eq:dftOfResolvent}
\end{multline}
where $\ket{\psi _{\rm in}}$ and $\ket{\psi _{\rm out}}$ are two many-body states. Here, we truncate the CP expansion (i.e., the second summation) utilizing the asymptotic form of $J_n(\pi \tau)$ and take $n_{\rm max} (\tau)> e \pi \tau /2 \approx 4\tau $  \cite{PhysRevC.79.044308}. The truncation error dramatically decreases with increasing $n_{\rm max} (\tau)$ \cite{10.1063/1.448136,PhysRevA.108.022422}. The Chebyshev moment $ \langle \psi _{\rm out} | T_n (H) | \psi _{\rm in} \rangle $ captures the many-body nature.

The resolvent has broad applications in solving the structure and dynamics of quantum many-body problems. As a first attempt, we focus on establishing the framework for quantum computing  many-fermion structure problems in this work. To date, this has not been discussed in the literature.

{\it Hamiltonian input scheme.--}
We introduce a novel and efficient {\it input scheme} for general second-quantized many-fermion Hamiltonian
\begin{equation}
	H= \sum _j \langle Q_j |H | P_j \rangle b^{\dagger} _{Q_j} b_{P_j} .
	\label{eq:HamiltonianLabeling}
\end{equation}
Here, for each {\it monomial} indexed by $j \in [0, \mathcal{D} -1 ]$, we denote $Q_j \mapsto \{ p_j , q_j, \cdots , r_j \} $ and $P_j \mapsto  \{u_j, v_j, \cdots , w_j  \}$, where we order the single-particle (SP) bases as $p_j < q_j < \cdots < r_j$ and $ u_j < v_j < \cdots < w_j $. The {\it few-body} matrix element is $ \langle Q_j |H | P_j \rangle \equiv \langle p_j , q_j, \cdots , r_j |H | u_j, v_j, \cdots , w_j  \rangle $. 
We also have $ b^{\dagger} _{Q_j} \equiv a^{\dagger}_{p_j} a^{\dagger}_{q_j} \cdots a^{\dagger}_{r_j} $ and $b_{P_j} \equiv a_{w_j} \cdots a_{v_j} a_{u_j}  $. Here the fermion operators act on the occupation mode of SP bases as $ a^{\dagger}\ket{0} = \ket{1},\ a\ket{1} =\ket{0},\ a^{\dagger}\ket{1} = a\ket{0} = 0 $; and they obey the anticommutation relations $	\{ a_p, a^{\dagger}_q \} = \delta _{p,q}, \ \{ a_p, a_q \} = \{a^{\dagger}_p, a^{\dagger}_q\} = 0$. 



This input model is based on the direct basis encoding scheme, which maps $N_{\rm sp}$ SP bases to the qubits in the system register ``$s$". A Fock state is then encoded as $ \ket{\mathcal{F}} = \ket{ b_0 }_0 \ket{b_1 }_1 \cdots  \ket{b_{N_{\rm sp}-1}}_{N_{\rm sp}-1}  $, where each qubit in $s$ corresponds to a specific SP basis, and $\ket{b_p}_p = \ket{0}$ or $\ket{1}$ denotes that the $p$th basis is vacant or occupied.

\begin{figure}[ht] 
	\centering
	\includegraphics[width=0.97\linewidth]{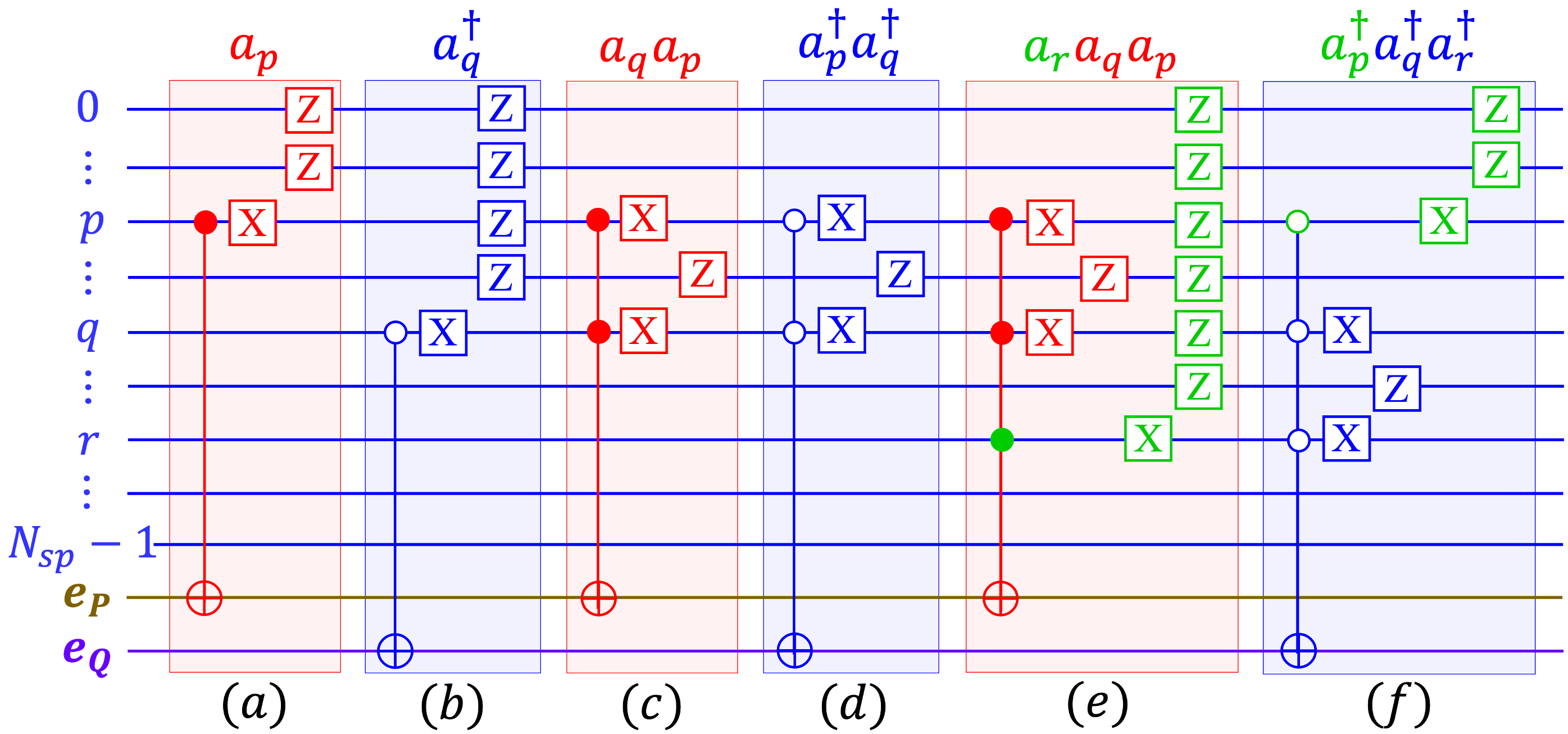}
	\caption{(color online) From left to right: the circuit representations of $a_p$, $a^{\dagger}_q$, $a_qa_p$, and $a^{\dagger}_p a^{\dagger}_q$, $a_r a_q a_p$, and $a^{\dagger}_p a^{\dagger}_q a^{\dagger}_r $ (with  $p<q<r$). The standard notations of quantum gates \cite{nielsen2010quantum} are adopted.
	} 
	\label{fig:combination1}
\end{figure}

We introduce systematic circuit representations for various combinations of fermion operators. 
These novel representations are reminiscent of Boolean masking \cite{helgaker2014molecular}. 
The fermion operators are represented by operations on qubit strings in the system register $s$ that encodes many-fermion states, where corresponding validation information is attached to each processed qubit string to distinguish physical operations from nonphysical ones.

We explain these circuit representations by illustrations in Fig.  \ref{fig:combination1}. In addition to the register $s$, we employ two ``$error$" registers, $e_P$ and $e_Q$ (both initialized as $\ket{1}$), to encode the validation information for the actions of annihilation and creation operators, respectively. For example, the circuit for $a_p$ operates as follows: (1) controlled by $ \ket{b_p}_p = \ket{1}$, we flip the state $\ket{y}_{e_P} = \ket{1}$ to $\ket{0}$; (2) we then apply a $X$ gate to $\ket{b_p}_p$, aiming to annihilate the occupation; (3) we employ a set of $Z$ gates on qubits with indices ranging from $0$ to $p-1$, whereby these $Z$ gates detect the phase factor due to anticommutation relations. To this end, we note that only a valid operation of $a_p$ on $\ket{b_p}_p$ results in the states $\ket{b_p} _p = \ket{0}$ and $\ket{y}_{e_P} = \ket{0} $.

We apply the same idea to design the circuit for $a^{\dagger}_q$, where, in this case, we flip $ \ket{y }_{e_Q} = \ket{1}$ to $\ket{0}$ controlled by $\ket{b_q}_q = \ket{0}$. Only a valid operation results in the states $ \ket{b_q}_q = \ket{1} $ and $\ket{y}_{e_Q}= \ket{0}$. 

With these elementary circuits, we can construct circuit representations of various combinations of fermion operators, as showcased in Fig. \ref{fig:combination1}(c)-(f). Additional optimizations are also applicable, e.g., cancellations of the $Z$ gates and simplifications of paired operators such as $a^{\dagger}_pa_p$. 

Our circuit representations of fermion operators are more intuitive and efficient than the JW and BK transformations, particularly in nuclear and particle physics applications where many-body interactions are present. Unlike the JW and BK transformations, which often require compiling each many-body operator into multiple Pauli strings, our design avoids this compilation overhead; this saving also reduces overhead in subsequent Hamiltonian encoding.


\begin{figure}[ht] 
	\centering
	\includegraphics[width=0.97\linewidth]{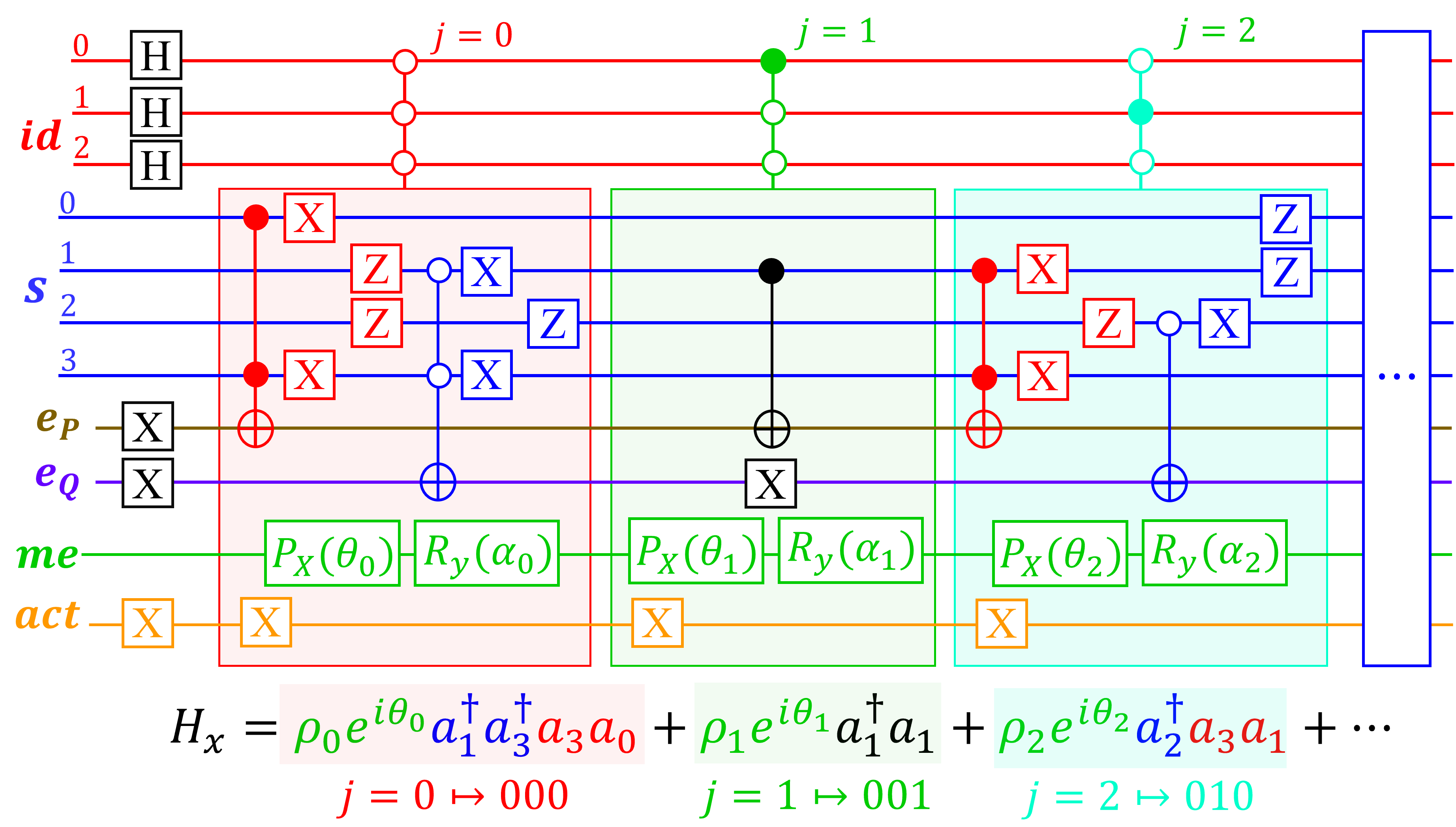}
	\caption{(color online) Illustration: the circuit of constructing the forward walk state based on the model Hamiltonian $H_{x}$. The register $s$ encodes Fock states. The $j \ (= 0, 1, 2)$ values index the monomials in $H_x$. Standard notations \cite{nielsen2010quantum} of elementary quantum gates are adopted except $P_X$ [Eq. \eqref{eq:PxGate}].  
	} 
	\label{fig:input_model_example}
\end{figure}




We can block-encode the second-quantized many-fermion Hamiltonian $H$ utilizing the circuit representations of various combinations of fermion operators. By leveraging the concept of Child's quantum walk \cite{childs2010relationship}, we construct the ``{\it forward}" walk state, the many-fermion state $\ket{\mathcal{F}}$, according to $H$ as 
\begin{align}
	 	\mathscr{T}_f \ket{\mathcal{F}}_s \otimes \ket{0} _a = & \frac{1}{\sqrt{\mathcal{B}}}  \sum _{j=0} ^{\mathcal{B}-1} \xi _{\mathcal{F},j}  e^{i \beta _j} \ket{j}_{id} \ket{\mathcal{F}_j} _{s} \nonumber \\
	 	& \times \ket{ y _{\mathcal{F},j}^P }_{e_P} \ket{ y _{\mathcal{F},j}^Q }_{e_Q} \ket{\rho _j}_{me} \ket{v_j}_{act} , 	
	 	\label{eq:forward_walk_state_maintext}
\end{align}
where $ \mathscr{T}_f $ defines the operation  to construct the walk state [see Supplementary Material (SM) for more details] and ``$a$" denotes all the ancilla registers. For explanatory purposes, we showcase the circuit for the walk-state construction in Fig. \ref{fig:input_model_example}. 

In Eq. \eqref{eq:forward_walk_state_maintext}, we identify each term in the summation with a specific $j$ value in the ``index" register $id$. Indeed, each term tracks the action of the $j$th monomial $\langle Q_j |H | P_j \rangle b^{\dagger} _{Q_j} b_{P_j}$ of $H$ on $\ket{\mathcal{F}}$ encoded in the register $s$, where the auxiliary information is recorded by the ancilla registers. $\xi _{\mathcal{F},j} $ records the actions of the $Z$ gates in our fermionic circuit representations and it takes the values of $\pm 1$. As mentioned above, we record the validation messages of sequential actions of $ b^{\dagger} _{Q_j} $ and $ b_{P_j}$ on $\ket{\mathcal{F}}$ by the states $\ket{y _{\mathcal{F},j}^Q }_{e_Q} $ and $\ket{y _{\mathcal{F},j}^P }_{e_P}$, respectively. While $y _{\mathcal{F},j}^Q$ and $y _{\mathcal{F},j}^P$ can be either 0 or 1, it is only when $y _{\mathcal{F},j}^P = y _{\mathcal{F},j}^Q = 0$ that $b^{\dagger} _{Q_j} b_{P_j} \ket{\mathcal{\mathcal{F}} } _s $ is valid and produces a physical many-fermion state $\ket{\mathcal{F}_j} _{s} $ with the corresponding factor $\xi _{\mathcal{F},j} $.

The matrix-element register ``$me$" records the scaled few-body matrix element $ \langle Q_j |H | P_j \rangle / \Xi $ by the state $e^{i \beta _j} \ket{\rho _j} _{me}$, with $  \ket{\rho _j} \equiv \rho _j \ket{0} +\sqrt{1-\rho ^2_j } \ket{1} $ and the {\it max norm} $\Xi \geq \max_j | \langle Q_j |H | P_j \rangle |$. Here the parameters are $\rho_j = | \langle Q_j |H | P_j \rangle | / \Xi  \leq 1$  and $\beta_j = \arg [\langle Q_j |H | P_j \rangle]$. In practice, the state $e^{i \beta _j} \ket{\rho _j} _{me}$ can be prepared by applying the modified phase gate $P_X(\beta _j )$ followed by a $R_y(\alpha _j)$ gate \cite{nielsen2010quantum} to the register state $ \ket{0} _{me} $, with
\begin{equation}
	P_X(\beta _j) \ket{0} = e^{i\beta _j} \ket{0}, \ P_X(\beta _j) \ket{1} = \ket{1},
	\label{eq:PxGate}
\end{equation}
and $ \alpha _j = 2\arccos \rho _j $.

The action register ``$act$" is introduced in practical calculations. Since we take $\mathcal{B} \geq \mathcal{D} $ in creating the walk state, $ \ket{v_j}_{act}$ (initialized as $\ket{1}$) serves to distinguish the {\it null} terms in the summation where the index $j$ does not correspond to any monomial in $H$. Those null terms are marked as $ v_j =1 $, while we have $v_j =0$ otherwise.

It is straightforward to construct the ``{\it backward}" walk state with the operation $ \mathscr{T}_b $ and the many-fermion state $\ket{\mathcal{G}}$ as
\begin{equation}
	\mathscr{T} _b \ket{\mathcal{G}}_s \otimes \ket{0} _a  = \frac{1}{\sqrt{\mathcal{B}}} \sum _{k=0} ^{\mathcal{B}-1} \ket{k}_{id} \ket{\mathcal{G}} _{s} \ket{0}_{e_P} \ket{0} _{e_Q} \ket{0}_{me} \ket{0}_{act} . \label{eq:backward_walk_state_maintext}
\end{equation}

These two walk states block-encode the scaled Hamiltonian $H'\equiv H/(\mathcal{B} \Xi )$ as (proof available in the SM)
\begin{equation}
	( \bra{\mathcal{G}} _s \otimes \bra{0} _a ) \mathscr{T} ^{\dagger} _b \mathscr{T}_f (\ket{\mathcal{F}}_s \otimes \ket{0} _a ) =  \frac{1}{\mathcal{B} \Xi} \langle \mathcal{G} | H | \mathcal{F} \rangle .
	\label{eq:Hamiltonian_input_scheme_maintext}
\end{equation}

There are at most $O(N^{2k}_{\rm sp})$ monomials in a general many-fermion Hamiltonian that includes up to $k$-body interactions.
Our scheme requires $\widetilde{O}(N^{2k+1}_{\rm sp})$ gates to input such a Hamiltonian, which represents the lowest rigorous upper bound on the gate cost for encoding a general $k$-body Hamiltonian. 
Further incorporation of symmetry principles can reduce this gate cost.

Compared to the conceptual Hamiltonian input scheme based on Child's quantum walk \cite{PhysRevLett.102.180501,childs2010relationship,berry2012black} and other input schemes \cite{Du:2023bpw,Liu:2024hmm,simon2025ladderoperatorblockencoding}, our scheme provides a practical circuit design, avoiding challenging oracle design, ancilla uncomputation, and quantum-classical communication for accessing Hamiltonian elements.

Compared to standard LCU \cite{Childs2012HamiltonianSU,PhysRevLett.114.090502}, our scheme eliminates the need for the ``Prepare oracle", which becomes increasingly difficult to design as matrix elements grow. It admits the max norm $\Xi$ for Hamiltonian scaling, which is smaller than the $\mathbb{L}^1$ norm in LCU \cite{RyanBabbushNJP2016}, making our approach more time-efficient for Hamiltonian simulations.

Utilizing our many-fermion Hamiltonian input scheme, we can block encode $T_n(H')$ following the standard protocol of Ref. \cite{PhysRevLett.118.010501, gilyen2019quantum, lin2022lecture}. 
The gate cost for encoding $T_n(H')$ is $\widetilde{O}(n\cdot N^{2k+1}_{\rm sp})$.
Specific circuit representations of the CPs are also available in the SM.

{\it Hybrid method.--}
We introduce a novel quantum-classical algorithm for the structure calculations of many-fermion systems such as those arising in nuclear and hadronic physics.
This algorithm is capable of computing the full bound spectrum and the $J$ values of the energy eigenstates. 

In particular, for our bound-state applications, we adopt the resolvent to compute the {\it spectral function} $ F_{\psi} (E_p) \equiv {\mathfrak{Re} }  \langle \psi | G(E_p) |  \psi  \rangle $ with the pivot state $\ket{\psi}$. $F_{\psi} (E_p)$ connects directly to the strength function \cite{PhysRevC.79.044308}, and its poles correspond to the eigenenergies.

\begin{figure*}[ht] 
	\centering
	\includegraphics[width=0.99\linewidth]{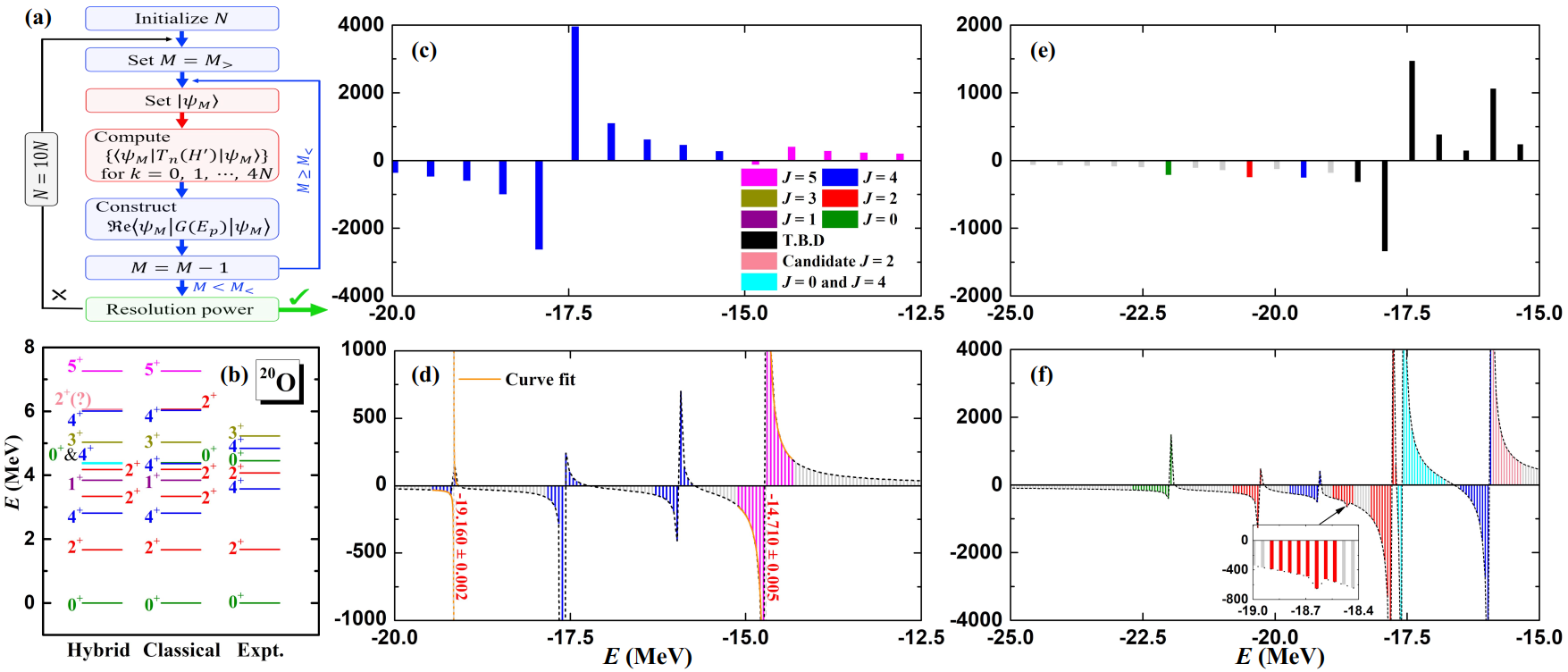}
	\caption{(color online) {\bf (a)} Algorithmic framework. {\bf (b}) Excitation energies and $J$ values of energy eigenstates of ${^{20}}O$ based on the hybrid scheme. These results are compared with classical calculations and experiments \cite{NNDC2022}. {\bf (c-f)} Spectral functions of ${^{20}}O$ as a function of energy $E$ [MeV]. {\bf (c)} and {\bf (d)} are computed with the pivot of the configuration $(1s_{\frac{1}{2},-\frac{1}{2}} ) (0d_{\frac{5}{2},\frac{1}{2}}) (0d_{\frac{5}{2},\frac{3}{2}} ) (0d_{\frac{5}{2},\frac{5}{2}} ) $ with $M=4$. {\bf (e)} and {\bf (f)} are computed with the pivot of the configuration $(1s_{\frac{1}{2},-\frac{1}{2}} ) (1s_{\frac{1}{2},\frac{1}{2}}) (0d_{\frac{5}{2},-\frac{1}{2}} ) (0d_{\frac{5}{2},\frac{1}{2}} ) $ with $M=0$. The number of grid points is $N=2000$ for {\bf (c)} and {\bf (e)}, while $N=20000$ for {\bf (d)} and {\bf (f)}. The peaks are fitted according to $f(x) = {a_1}/{(x - E_x + 10^{-6})} + {a_2}/{(x - E_x + 10^{-6})^2} + \eta$ with $a_1$, $a_2$, $E_x$ and $\eta$ being the unknowns [see examples in panel (d)].
	} 
	\label{fig:combined_pic}
\end{figure*}

The algorithmic workflow is sketched in Fig. \ref{fig:combined_pic}(a). We compute the Chebyshev moments $\{ \langle \psi |T_n(H') | \psi \rangle \}$ based on our Hamiltonian input scheme. These moments are input to the classical computer to construct $ F_{\psi} (E_p) $, from which we resolve the spectrum. As our input scheme preserves the symmetry of the Hamiltonian, it enables the so-called ``$M$-scheme" calculations. Namely, if we input the pivot to be of a specific projection $M$ of $J$, then only states of $J\geq |M|$ emerge in the resultant spectrum.

This algorithm admits a two-fold scan scheme: the cascading-$M$ scan and the resolution scan. 
With a sufficient resolution scale (set by the number of grid points $N$), only those energy eigenstates with $J\geq M \geq 1$ appear in the spectrum if we compute using the pivot $\psi _M$ of the projection $M$.
If additional states emerge when computing  $ F_{\psi} (E_p) $ with the pivot $\psi _{M-1}$ of the projection $M-1$, the $J$ value of these emerging states are $J=M-1$. Following this idea, if we compute a set of $ F_{\psi} (E_p) $ based on pivots with projections in the range $M \in [M_< , M_>] $ (with $M_> > M_< \geq 0$), we can distinguish those eigenstates with $J = M_> -1, \cdots , J= M_<$. 

Finer resolutions may be needed to distinguish closely spaced states. In this case, we improve the resolution by increasing the number of grid points $N$, thereby incorporating more Chebyshev moments. In practical calculations, we can start with a moderate $N$ value for a rough scan. Finer scans with larger $N$ values follow if one wants to resolve even more closely spaced energy eigenstates.

We evaluate the total number of Chebyshev moments required to construct $ F_{\psi} (E_p) $ is $O(\mathcal{B} \Xi /\Delta )$, where $\Delta $ denotes the spectral gap to be resolved between two neighboring states of interest (in, e.g., a low-lying spectrum). 

{\it Realistic calculations.--}
We apply our hybrid method to compute the structure of ${^{20}}O$ with a non-relativistic two-body Hamiltonian in a 3D harmonic oscillator (HO) basis. Had we selected a BLFQ example for a hadron, we would employ a 2D HO basis for the transverse motion in light-front coordinates. In this work, we retain the SP bases in the $0d_{5/2}1s_{1/2}$ shell and exclude those in the $0d_{3/2}$ shell. 
We implement the realistic nuclear interaction from Ref. \cite{Shin:2024zpe}, where we adopt a basis transformation to obtain the few-body Hamiltonian matrix elements in our SP basis set. A total of 110 matrix elements are retained 
(see Table \ref{tab:few_body_matrix_elements} in the SM). The same techniques can be applied to no-core shell model \cite{Navratil:2000ww,Navratil:2000gs,Barrett:2013nh} calculations in future research.

We evaluate the Chebyshev moments using the IBM Qiskit \cite{Qiskit} Statevector simulator in noiseless mode; these moments are cross-checked against corresponding classical calculations. In future applications on real quantum hardware, these moments can be evaluated using standard techniques such as the Hadamard test \cite{nielsen2010quantum}.

We apply our two-fold scan scheme to obtain the $J$ values and eigenenergies of the states utilizing simple single-configuration pivots (note that alternative single-configuration pivots may be needed to unconver missing states; additional comments are available in the SM). By applying the cascading-$M$ scan at a coarse resolution scale [Fig. \ref{fig:combined_pic}(c) and (e)], we can obtain a rough estimation of the eigenenergies and $J$ values of the states that are well-separated from others. With an improved resolution, more spectral information is resolved [Fig. \ref{fig:combined_pic}(d) and (f)]. In Fig. \ref{fig:combined_pic}(f), we note regions (marked in cyan and light pink) where finer resolutions are necessary to resolve states with small spectral gaps. Nevertheless, a coarse resolution scale would be sufficient for the low-lying states with larger spectral gaps.


The excitation spectrum of ${^{20}}O$ is presented in Fig. \ref{fig:combined_pic}(b) with the corresponding $J$ values. Our results from the hybrid method agree with the classical results. We expect better agreement with experiments \cite{NNDC2022} when the SP states of the $0d_{3/2}$ orbit are included.

{\it Summary.--}
We propose a novel quantum-classical framework for {\it ab initio} many-fermion structure calculations. This hybrid method is based on an efficient input scheme for general second-quantized many-fermion Hamiltonians. Using this scheme, we quantum compute the Chebyshev moments and input them into a classical computer to calculate the resolvent, from which we obtain the Hamiltonian spectral function. We introduce a two-fold scan scheme to resolve both the energies and $J$ values of the eigenstates from the spectral function. We apply this hybrid method to compute, for the first time, the structure of ${^{20}}O$ based on a realistic strong-interaction Hamiltonian, where we resolve the energies and $J$ values of most eigenstates.

Our approach is directly applicable to hadron spectroscopy \cite{Vary:2009gt,Qian:2020utg,Kreshchuk:2020dla,Kreshchuk:2020aiq,Du:2019ips,Du:2023bpw}. 
We also  envision our framework being applicable to CI calculations in fields such as quantum chemistry \cite{Cao_2019, McArdle_2020}, condensed matter physics \cite{Yoshioka:2022rej, PRXQuantum.2.030307},  and field theories \cite{Bauer:2023qgm, bauer2022quantum, Kreshchuk:2020dla, Kreshchuk:2020aiq}. 
Our method easily generalizes to evaluate static and reaction observables using the well-known response theory \cite{stefanucci2025nonequilibrium,fetter2003quantum,Carlson:2014vla}.
Further developments aim to improve circuit designs by utilizing, e.g., quantum machine learning \cite{Biamonte_2017, Wang_2024}, for calculations on NISQ hardware \cite{Preskill2018quantumcomputingin}. 

{\it Acknowledgements.--}
We acknowledge fruitful discussions with Peter Love, Pieter Maris, and Andrey M. Shirokov. This work was supported in part by US DOE Grant DE-SC0023707 under the Office of Nuclear Physics Quantum Horizons program for the “Nuclei and Hadrons with Quantum computers (NuHaQ)” project. This project was also supported in part by NSF Grant No. 2435255 (NQVL-QSTD: Q-BLUE).

\bibliography{apssamp}

\newpage


\clearpage
\onecolumngrid
\begin{center}
	\textbf{\large Supplementary Material: \\
Ab initio many-fermion structure calculations on a quantum computer
}

\renewcommand{\thefootnote}{\fnsymbol{footnote}}
\setcounter{footnote}{0} 

{\small Weijie Du$^{1,2}$, Yangguang Yang$^{1}$, Zixin Liu$^{1}$,\footnote{zixinliu@gdlhz.ac.cn} Chao Yang$^{3}$, and James P. Vary$^{2}$}\\[5pt]
{\small $^{1}$Institute of Modern Physics, Chinese Academy of Sciences, Lanzhou 730000, China}\\
{\small $^{2}$ Department of Physics and Astronomy, Iowa State University, Ames, Iowa 50010, USA}\\
{\small $^{3}$Applied Mathematics and Computational Research Division, Lawrence Berkeley National Laboratory, Berkeley, California 94720, USA}\\[10pt]
\end{center}



\setcounter{section}{0}
\renewcommand{\thesection}{S\arabic{section}}

\setcounter{section}{0}
\renewcommand{\thesection}{S\arabic{section}}

\setcounter{figure}{0}
\renewcommand{\thefigure}{S\arabic{figure}}

\setcounter{table}{0}
\renewcommand{\thetable}{S\arabic{table}}

\section{Walk states construction and Hamiltonian input scheme}

We first construct the {\it forward} walk state. Provided the Fock state $\ket{\mathcal{F}}$, we prepare the state of multiple quantum registers as
\begin{equation}
	\ket{\Phi _0} = \ket{0}_{id} \ket{\mathcal{F}} _{s} \ket{1}_{e_P} \ket{1} _{e_Q} \ket{0}_{me} \ket{1}_{act} .
\end{equation}
Here the index register ``$id$" is initialized as $\ket{0}$. The qubits in the system register ``{\it $s$}" encode the Fock state $\ket{\mathcal{F}}$ in terms of the binary string $\{b_0b_1\cdots b_{N_{\rm sp}-1} \}$. The single-qubit registers ``$e_P$", ``$e_Q$" and ``$act$" are initialized as $\ket{1}$. The single-qubit register ``$ me $"  is initialized as $\ket{0}$. The roles of these registers will be clear in the following text.

We first apply the diffusion operator to $id$ register and $\ket{\Phi _0}$ becomes
\begin{equation}
	\ket{\Phi _1} =  \frac{1}{\sqrt{\mathcal{B}}} \sum _{j=0} ^{\mathcal{B}-1} \ket{j}_{id} \ket{\mathcal{F}} _{s} \ket{1}_{e_P} \ket{1} _{e_Q} \ket{0}_{me} \ket{1}_{act} ,
\end{equation}
where $\mathcal{B} \geq \mathcal{D}$ with $\mathcal{D}$ denoting the total number of monomials in $H$ [Eq. \eqref{eq:HamiltonianLabeling} in the main text]. Note that the diffusion operator can produce more $j$ values than the total number of monomials in $H$.
We label each monomial in $H$ with a specific index $j $. The $j$th monomial takes the form of $\langle Q_j |H | P_j \rangle b^{\dagger} _{Q_j} b_{P_j}$. Then, we implement the circuit operations of $b_{P_j}$ and $b^{\dagger} _{Q_j}$ controlled by $j$ encoded in the index register. With these implementations, we have 
\begin{equation}
	\ket{j}_{id} \ket{\mathcal{F}} _{s} \ket{1}_{e_P} \ket{1} _{e_Q} \ket{0}_{me} \ket{1}_{act} \rightarrow \ket{j}_{id} \xi _{\mathcal{F},j} \ket{\mathcal{F}_j} _{s} \ket{ y _{\mathcal{F},j}^P }_{e_P} \ket{ y _{\mathcal{F},j}^Q }_{e_Q} e^{i \beta _j} \ket{\rho _j}_{me} \ket{0}_{act},
	\label{eq:sample_j}
\end{equation}
where we have $\xi _{\mathcal{F},j} \ket{\mathcal{F}_j} _{s} = b^{\dagger} _{Q_j} b_{P_j} \ket{\mathcal{\mathcal{F}} } _s$. The factor $ \xi _{\mathcal{F},j} $ takes the value of $\pm 1$, which is obtained from the sequences of Pauli-$Z$ gates in the circuit representation of the fermion operators acting on respective sequences of qubits that encode the occupation status of relevant SP bases. $\ket{\mathcal{F}_j} _{s} $ is a binary string. 
While $ b^{\dagger} _{Q_j} b_{P_j} \ket{\mathcal{\mathcal{F}} } _s $ should be understood formally, their circuit execution may produce binary strings $\ket{\mathcal{F}_j} _{s}$ that is nonphysical.
As mentioned above, we record the validation message of the sequential actions of $ b^{\dagger} _{Q_j} $ and $ b_{P_j}$ on $\ket{\mathcal{F}}$ by the states $\ket{y _{\mathcal{F},j}^Q }_{e_Q} $ and $\ket{y _{\mathcal{F},j}^P }_{e_P}$, respectively. While both $y _{\mathcal{F},j}^Q$ and $y _{\mathcal{F},j}^P$ can take the value of 0 or 1, it is only when $y _{\mathcal{F},j}^P = y _{\mathcal{F},j}^Q = 0$ that $b^{\dagger} _{Q_j} b_{P_j} \ket{\mathcal{\mathcal{F}} } _s $ produces a physical many-fermion state $\ket{\mathcal{F}_j} _{s} $ with the corresponding phase factor $\xi _{\mathcal{F},j} $.

We operate on the single-qubit register $me$ in Eq. \eqref{eq:sample_j} according to the scaled {\it few-body} matrix element $\langle Q_j |H | P_j \rangle /\Xi$ with $\Xi \geq \max_j | \langle Q_j |H | P_j \rangle |$. The operation is 
\begin{equation}
	\ket{0}_{me} \rightarrow e^{i \beta _j} \ket{\rho _j} _{me},
	\label{eq:encoding_me}
\end{equation}
where $\ket{\rho _j} \equiv \rho _j \ket{0} +\sqrt{1-\rho ^2_j } \ket{1} $. The parameters $\rho _j$ and $\beta _j $ are determined as 
\begin{equation}
	\rho _j e^{i \beta _j} =  \langle Q_j |H | P_j \rangle / \Xi ,
	\label{eq:scaled_few_body_Elem}
\end{equation}
where $\rho_j = | \langle Q_j |H | P_j \rangle | / \Xi  \leq 1$ with  and $\beta_j = \arg [\langle Q_j |H | P_j \rangle]$.
The operation [Eq. \eqref{eq:encoding_me}] can be achieved by applying the gate $P_X(\beta _j )$ and sequentially by a $R_y(\alpha _j)$ gate \cite{nielsen2010quantum} to the register $me$, where
\begin{equation}
	P_X(\beta _j) \ket{0} = e^{i\beta _j} \ket{0}, \ P_X(\beta _j) \ket{1} = \ket{1}, \ \alpha _j = 2\arccos \rho _j .
\end{equation}

In Eq. \eqref{eq:sample_j}, we also flip the register state $ \ket{1}_{act}  $ to $\ket{0}_{act} $ by a Pauli-$X$ gate as the $j$ value indeed corresponds to a monomial. 

Following the operations illustrated by Eq. \eqref{eq:sample_j}, we have the forward walk state
\begin{equation}
	\ket{\Phi _2} = \frac{1}{\sqrt{\mathcal{B}}} \sum _{j=0} ^{\mathcal{B}-1} \xi _{\mathcal{F},j}  e^{i \beta _j} \ket{j}_{id} \ket{\mathcal{F}_j} _{s} \ket{ y _{\mathcal{F},j}^P }_{e_P} \ket{ y _{\mathcal{F},j}^Q }_{e_Q} \ket{\rho _j}_{me} \ket{v_j}_{act}.
	\label{eq:forward_state}
\end{equation}
Here we denote the state of $act$ as $\ket{v_j}_{act}$ with $v_j = 0$ or $1$. The state $\ket{v_j}_{act}$ helps to distinguish the true contribution from the Hamiltonian monomials to the Fock state $\ket{\mathcal{F}}$, noticing that more $j$ values can be produced by the diffusion operator than the number of Hamiltonian monomials in practical calculations (recall we have $\mathcal{B} \geq \mathcal{D} $). 
If the $ j $ value does correspond to a specific Hamiltonian monomial, we have $v_j = 0$, and $v_j $ is 1 otherwise. In Fig. \ref{fig:input_model_example}, we present an illustration of constructing the forward walk state based on a model Hamiltonian.

The {\it backward} walk state $\ket{\Psi _1} $ is prepared as follows. Starting from the Fock state $\ket{\mathcal{G}}$, we initialize the multi-register state
\begin{equation}
	\ket{\Psi _0} = \ket{0}_{id} \ket{\mathcal{G}} _{s} \ket{0}_{e_P} \ket{0} _{e_Q} \ket{0}_{me} \ket{0}_{act},
\end{equation}
where the registers $e_P$, $e_Q$, $me$, and $act$ are all initialized as $\ket{0}$, which is to be compared with the initialization of $\ket{\Phi_0}$. We then apply the diffusion operator to the register $id$, and $\ket{\Psi _0}$ becomes
\begin{equation}
	\ket{\Psi _1} = \frac{1}{\sqrt{\mathcal{B}}} \sum _{k=0} ^{\mathcal{B}-1} \ket{k}_{id} \ket{\mathcal{G}} _{s} \ket{0}_{e_P} \ket{0} _{e_Q} \ket{0}_{me} \ket{0}_{act} .
	\label{eq:backward_state}
\end{equation}

The forward and backward walk states block encode the Hamiltonian of the many-fermion system. Indeed, their inner product is
\begin{align}
	\langle \Psi _1 | \Phi _2 \rangle 
	=& \frac{1}{\mathcal{B}} \sum _{j=0} ^{\mathcal{B}} \sum _{k=0} ^{\mathcal{B}} \xi _{\mathcal{F},j}  e^{i \beta _j}  \langle k | j \rangle _{id} \langle \mathcal{G} | \mathcal{F}_j \rangle _{s} \langle 0 | y _{\mathcal{F},j}^P \rangle _{e_P} \langle 0 | y _{\mathcal{F},j}^Q \rangle_{e_Q} \langle 0 | \rho _j \rangle _{me} \langle 0 | v_j \rangle _{act} \\
	= & \frac{1}{\mathcal{B}} \sum _{j=0} ^{\mathcal{B}} \xi _{\mathcal{F},j}  e^{i \beta _j}   \langle \mathcal{G} | \mathcal{F}_j \rangle _{s} \langle 0 | y _{\mathcal{F},j}^P \rangle _{e_P} \langle 0 | y _{\mathcal{F},j}^Q \rangle_{e_Q} \langle 0 | \rho _j \rangle _{me} \langle 0 | v_j \rangle _{act} ,
\end{align}
which correctly produces the many-fermion matrix element of the scaled Hamiltonian $H' = H /(\mathcal{B} \Xi) $ based on the Fock states $\ket{\mathcal{F}}$ and $\ket{\mathcal{G}}$ as well as scaled few-body matrix elements $\{ \rho _j e^{i \beta _j} \}$ [Eq. \eqref{eq:scaled_few_body_Elem}].  

We can present $\langle \Psi _1 | \Phi _2 \rangle $ in terms of the standard form of Hamiltonian block encoding. In particular, we can write the walks states based on the isometries $\mathscr{T}_f$ and $\mathscr{T}_b$ as
\begin{equation}
	\ket{\Phi _2} = \mathscr{T}_f \ket{\mathcal{F}}_s \otimes \ket{0} _a , \ \ket{\Psi _1 } = \mathscr{T} _b \ket{\mathcal{G}}_s \otimes \ket{0} _a .
\end{equation}
Here, the ancilla registers are denoted as ``$a$", which include all the registers except the system register $s$. Then, the Hamiltonian block encoding scheme can be presented as
\begin{equation}
	( \bra{\mathcal{G}} _s \otimes \bra{0} _a ) \mathscr{T} ^{\dagger} _b \mathscr{T}_f (\ket{\mathcal{F}}_s \otimes \ket{0} _a ) =  \frac{1}{\mathcal{B} \Xi} \langle \mathcal{G} | H | \mathcal{F} \rangle .
	\label{eq:Hamiltonian_input_scheme_SM}
\end{equation}

A few comments on the above Hamiltonian input scheme are as follows. This scheme is non-Hermitian, i.e., $\mathscr{T} ^{\dagger} _b \mathscr{T}_f \neq \big( \mathscr{T} ^{\dagger} _b \mathscr{T}_f \big)^{\dagger}$. While the standard quantum-walk-based input scheme, designed according to Childs’ quantum walk, operates in the first-quantization framework and accesses Hamiltonian matrix elements through communication with a precomputed classical database based on row and column indices \cite{PhysRevLett.102.180501,childs2010relationship,berry2012black}, our input scheme works in the second-quantization framework, dynamically computing many-body matrix elements of the scaled Hamiltonian using Fock states, without requiring data exchange with a precomputed classical database.

Meanwhile, our block encoding scheme eliminates the need for uncomputation, which can be costly in typical quantum algorithms \cite{PhysRevLett.102.180501,childs2010relationship,berry2012black}. Additionally, while the standard walk-state-based input scheme cannot provide a practical circuit design for general Hamiltonians due to challenges in quantum-classical communication, our scheme offers a viable circuit design.

\section{Circuit representation of the Chebyshev polynomials}

\begin{figure}[ht] 
	\centering
	\includegraphics[width=0.47\linewidth]{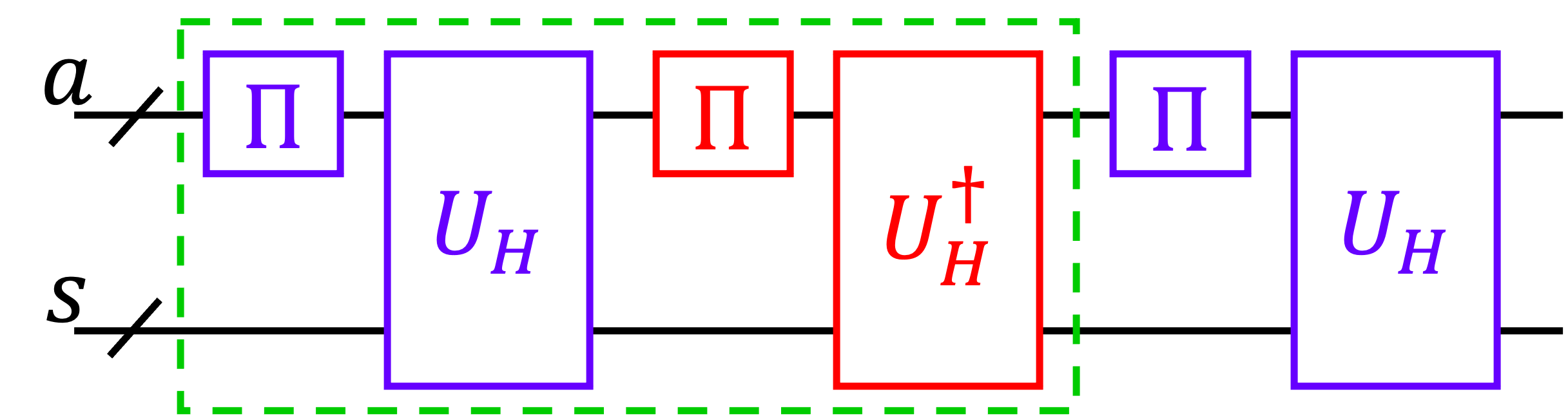}
	\caption{(color online) Illustration: Circuit for inputting $T_3(H')$ (figure adapted from  \cite{Du:2024ixj}). Here, $s$ and $a$ denote the system register and the ancilla register, respectively. The slashed line denotes a register that may contain multiple qubits. The circuit enclosed by the dashed green box block-encodes $T_2(H')$.  
	} 
	\label{fig:qubitization_pic}
\end{figure}

With the Hamiltonian input scheme [Eq. \eqref{eq:Hamiltonian_input_scheme_SM}], we can input the Chebyshev polynomials of the first kind  \cite{lin2022lecture,Du:2024zvr,Du:2024ixj} as
\begin{align}
	\langle \mathcal{G} | T_{2k+1} (H') | \mathcal{F} \rangle =& ( \bra{\mathcal{G}}_s \otimes \bra{0} _a ) \big[ U_H \Pi \big( U_H^{\dagger} \Pi U_H \Pi \big) ^k  \big] \big( \ket{\mathcal{F}}_s \otimes \ket{0}_a   \big), \\
	\langle \mathcal{G} | T_{2k} (H') | \mathcal{F} \rangle =& ( \bra{\mathcal{G}}_s \otimes \bra{0} _a ) \big( U_H^{\dagger} \Pi U_H \Pi \big) ^k  \big( \ket{\mathcal{F}}_s \otimes \ket{0}_a   \big),
\end{align}
with $k=0, \ 1, \ 2, \cdots $. Here, the reflection operator $\Pi \equiv (2 \ket{0}_a \bra{0}_a - \mathds{I} _a) \otimes \mathds{I}_s$ produces the reflection around $\ket{0}_a$ in the auxiliary space. We also define $U_H \equiv \mathscr{T}^{\dagger} _b \mathscr{T}_f$. Therefore, the quantum circuit for inputting $T_n(H')$ is achieved by alternating applications of the building blocks $U_H \Pi$ and $ U_H ^{\dagger} \Pi $ \cite{lin2022lecture}, as illustrated in Fig. \ref{fig:qubitization_pic}.

\section{Realistic Hamiltonian for the nuclear many-body calculation}

We compute the eigenenergies and the corresponding total angular momenta of the eigenstates of $ ^{20}O$ in this work. We 
adopt a subspace of the $sd$-valence shell. In particular, we retain the single-particle (SP) basis states in the $0d_{5/2}1s_{1/2}$ orbits, where we exclude those SP bases in the $0d_{3/2}$ orbit. The adopted SP bases are listed in Table \ref{tab:SP_basis}.

\begin{table}[ht]
	\centering
	\caption{The SP bases in the $0d_{5/2}1s_{1/2}$ valence shell. The quantum numbers of each SP state are presented, whereas the spin and isospin of the neutrons are understood to be $1/2$, while the isospin projection is $-1/2$. Each SP basis is indexed and mapped to a specific qubit.}
	\begin{tabular}{cccccccc} 
		\hline \hline
		& SP basis (qubit) & $n$ & $l$ & $2j$ & $2m$ &  \\ \hline
		\multirow{2}{*}{$1s_{1/2}$} & 0              & 1   & 0   & 1    & $-1$    \\
		& 1                & 1   & 0   & 1    & $+1$    \\ \hline
		\multirow{6}{*}{$0d_{5/2}$} & 2                & 0   & 2   & 5    & $-5$    \\
		& 3               & 0   & 2   & 5    & $-3$    \\
		& 4               & 0   & 2   & 5    & $-1$    \\
		& 5               & 0   & 2   & 5    & $+1$    \\
		& 6              & 0   & 2   & 5    & $+3$    \\
		& 7               & 0   & 2   & 5    & $+5$    \\ \hline
		\hline \hline
	\end{tabular} 
	\label{tab:SP_basis}
\end{table}

The Hamiltonian of the many-neutron system is 
\begin{equation}
	H = E_{\rm core} + \sum _{i=0} ^7 \varepsilon _i a^{\dagger}_i a_i + \sum _{p<q} ^7 \sum _{r<s} ^7 \langle p q | V | r s \rangle a^{\dagger}_p a^{\dagger}_q a _s a _r ,
	\label{eq:hamiltonian_expression_SM}
\end{equation}
where we adopt the phenomenologically successful choices of Ref. \cite{Shin:2024zpe} and set the SP energy $\varepsilon _i = -3.2079 $ MeV for $i =0, 1$ and $\varepsilon _i = -3.9257$ MeV for $i \in [2, 7]$. We set the core energy to be $E_{\rm core}=0$ for computing the excitation spectrum in our work.
The two-body interaction matrix elements are derived from the effective $sd$-shell NN interaction \cite{Shin:2024zpe}.
In Table \ref{tab:sd_nn_effective_interaction}, we list these interaction matrix elements in the coupled-$JT$ basis. 

\begin{table}[ht]
	\centering
	\caption{Effective $sd$-shell NN interaction matrix elements in the coupled $JT$-basis.These elements are quoted from Table S1 in Ref. \cite{Shin:2024zpe} with $J= 0,\ 1,\ 2,\ 3, \ 4$ and the isospin $T=1$.}
	\begin{tabular}{cc|cc|c|c||c}
		\hline \hline 
		$2j_a$ & $2 j_b$ & $2 j_c$ & $2j_d$ & $J$ & $T$ & $ V $ \\ \hline 
		1 & 1 & 1 & 1 & 0 & 1   &  -1.910 \\
		1 & 1 & 5 & 5 &  0 & 1 & -1.564 \\
		5 & 5 & 5  & 5 & 0 & 1 &  -2.439 \\ \hline 
		1 & 5 & 1 & 5 & 2 & 1 & -1.032 \\ 
		1 &  5 & 5 &  5 & 2 & 1 & -0.544 \\
		5 & 5 & 5 & 5 & 2 & 1  & -1.119 \\ \hline 
		1 & 5 & 1 & 5 & 3 & 1 & 0.751 \\ \hline
		5 & 5 & 5 & 5 & 4 & 1 & -0.152  \\
		\hline  \hline 
	\end{tabular}
	\label{tab:sd_nn_effective_interaction}
\end{table}

We can implement Eq. (8.24) in Ref. \cite{suhonen2007nucleons} to transform the interaction matrix elements from the coupled-$JT$ basis to the elected SP basis. 
These transformed two-body interaction matrix elements are listed in Table \ref{tab:few_body_matrix_elements} (with indices from 8 to 109). It is noteworthy that an additional factor $(18/A)^{1/3}$ (with $A=20$) is necessary to scale the interaction matrix elements provided in Table S3 for computing the spectrum of ${^{20}}O$. Indeed, these interaction matrix elements are also applicable to the spectral calculations of additional Oxygen isotopes using the corresponding scaling.

\begin{table*}
	\centering
	\caption{The matrix elements in the SP basis. The indices $p,\ q,\ r$, and $s$ label the SP bases listed in Table \ref{tab:SP_basis}. The first 8 entries are the SP energies \cite{Shin:2024zpe}. The rest entries (with index from 8 to 109) are the two-body interaction matrix elements in the SP basis.
	}
	\label{tab:few_body_matrix_elements}
	\begin{tabular}%
		{@{}c@{\hspace{4mm}}c@{\hspace{4mm}}c@{\hspace{4mm}}c@{\hspace{4mm}}c@{\hspace{4mm}}r@{\hspace{6mm}}c@{\hspace{4mm}}c@{\hspace{4mm}}c@{\hspace{4mm}}c@{\hspace{4mm}}c@{\hspace{4mm}}r@{\hspace{6mm}}c@{\hspace{4mm}}c@{\hspace{4mm}}c@{\hspace{4mm}}c@{\hspace{4mm}}c@{\hspace{4mm}}r@{}l}
		
		\toprule
		\midrule
		$\rm {index}$ & $p$ &  $q$ & $r$ & $s$ & $ \langle p q | H | r s \rangle $ & $\rm {index}$ & $p$ &  $q$ & $r$ & $s$ & $ \langle p q | H | r s \rangle $ & $\rm {index}$ & $p$ &  $q$ & $r$ & $s$ & $ \langle p q | H | r s \rangle $  \\
		
		\midrule
		
		$  0$&$ 0$&$ -$&$ 0$&$  -$&$ -3.2079  $&$  37$&$ 4$&$ 5$&$ 7$&$  2$&$ -0.30186 $&$  74$&$ 6$&$ 7$&$ 7$&$  6$&$ -0.15200 $ \\
		$  1$&$ 1$&$ -$&$ 1$&$  -$&$ -3.2079  $&$  38$&$ 4$&$ 5$&$ 6$&$  3$&$  0.85443 $&$  75$&$ 0$&$ 4$&$ 4$&$  0$&$  0.15667 $ \\
		$  2$&$ 2$&$ -$&$ 2$&$  -$&$ -3.9257  $&$  39$&$ 4$&$ 5$&$ 5$&$  4$&$ -1.28270 $&$  76$&$ 0$&$ 4$&$ 3$&$  1$&$  0.84051 $ \\
		$  3$&$ 3$&$ -$&$ 3$&$  -$&$ -3.9257  $&$  40$&$ 0$&$ 6$&$ 6$&$  0$&$ -0.43767 $&$  77$&$ 0$&$ 4$&$ 6$&$  2$&$ -0.26545 $ \\
		$  4$&$ 4$&$ -$&$ 4$&$  -$&$ -3.9257  $&$  41$&$ 0$&$ 6$&$ 5$&$  1$&$  0.84051 $&$  78$&$ 0$&$ 4$&$ 5$&$  3$&$  0.16788 $ \\
		$  5$&$ 5$&$ -$&$ 5$&$  -$&$ -3.9257  $&$  42$&$ 0$&$ 6$&$ 7$&$  3$&$ -0.37540 $&$  79$&$ 1$&$ 3$&$ 4$&$  0$&$  0.84051 $ \\
		$  6$&$ 6$&$ -$&$ 6$&$  -$&$ -3.9257  $&$  43$&$ 0$&$ 6$&$ 6$&$  4$&$  0.23742 $&$  80$&$ 1$&$ 3$&$ 3$&$  1$&$ -0.43767 $ \\
		$  7$&$ 7$&$ -$&$ 7$&$  -$&$ -3.9257  $&$  44$&$ 1$&$ 5$&$ 6$&$  0$&$  0.84051 $&$  81$&$ 1$&$ 3$&$ 6$&$  2$&$  0.37540 $ \\
		$  8$&$ 0$&$ 1$&$ 1$&$  0$&$ -1.9100  $&$  45$&$ 1$&$ 5$&$ 5$&$  1$&$  0.15667 $&$  82$&$ 1$&$ 3$&$ 5$&$  3$&$ -0.23742 $ \\
		$  9$&$ 0$&$ 1$&$ 7$&$  2$&$ -0.90298 $&$  46$&$ 1$&$ 5$&$ 7$&$  3$&$  0.26545 $&$  83$&$ 2$&$ 6$&$ 4$&$  0$&$ -0.26545 $ \\
		$ 10$&$ 0$&$ 1$&$ 6$&$  3$&$  0.90298 $&$  47$&$ 1$&$ 5$&$ 6$&$  4$&$ -0.16788 $&$  84$&$ 2$&$ 6$&$ 3$&$  1$&$  0.37540 $ \\
		$ 11$&$ 0$&$ 1$&$ 5$&$  4$&$ -0.90298 $&$  48$&$ 3$&$ 7$&$ 6$&$  0$&$ -0.37540 $&$  85$&$ 2$&$ 6$&$ 6$&$  2$&$ -0.84271 $ \\
		$ 12$&$ 0$&$ 5$&$ 5$&$  0$&$ -0.14050 $&$  49$&$ 3$&$ 7$&$ 5$&$  1$&$  0.26545 $&$  86$&$ 2$&$ 6$&$ 5$&$  3$&$  0.43685 $ \\
		$ 13$&$ 0$&$ 5$&$ 4$&$  1$&$  0.89150 $&$  50$&$ 3$&$ 7$&$ 7$&$  3$&$ -0.84271 $&$  87$&$ 3$&$ 5$&$ 4$&$  0$&$  0.16788 $ \\
		$ 14$&$ 0$&$ 5$&$ 7$&$  2$&$ -0.29678 $&$  51$&$ 3$&$ 7$&$ 6$&$  4$&$  0.43685 $&$  88$&$ 3$&$ 5$&$ 3$&$  1$&$ -0.23742 $ \\
		$ 15$&$ 0$&$ 5$&$ 6$&$  3$&$ -0.05936 $&$  52$&$ 4$&$ 6$&$ 6$&$  0$&$  0.23742 $&$  89$&$ 3$&$ 5$&$ 6$&$  2$&$  0.43685 $ \\
		$ 16$&$ 0$&$ 5$&$ 5$&$  4$&$  0.23742 $&$  53$&$ 4$&$ 6$&$ 5$&$  1$&$ -0.16788 $&$  90$&$ 3$&$ 5$&$ 5$&$  3$&$ -0.42829 $ \\
		$ 17$&$ 1$&$ 4$&$ 5$&$  0$&$  0.89150 $&$  54$&$ 4$&$ 6$&$ 7$&$  3$&$  0.43685 $&$  91$&$ 0$&$ 3$&$ 3$&$  0$&$  0.45383 $ \\
		$ 18$&$ 1$&$ 4$&$ 4$&$  1$&$ -0.14050 $&$  55$&$ 4$&$ 6$&$ 6$&$  4$&$ -0.42829 $&$  92$&$ 0$&$ 3$&$ 2$&$  1$&$  0.66449 $ \\
		$ 19$&$ 1$&$ 4$&$ 7$&$  2$&$  0.29678 $&$  56$&$ 0$&$ 7$&$ 7$&$  0$&$ -0.73483 $&$  93$&$ 0$&$ 3$&$ 5$&$  2$&$ -0.13272 $ \\
		$ 20$&$ 1$&$ 4$&$ 6$&$  3$&$  0.05936 $&$  57$&$ 0$&$ 7$&$ 6$&$  1$&$  0.66449 $&$  94$&$ 0$&$ 3$&$ 4$&$  3$&$  0.17807 $ \\
		$ 21$&$ 1$&$ 4$&$ 5$&$  4$&$ -0.23742 $&$  58$&$ 0$&$ 7$&$ 7$&$  4$&$ -0.29678 $&$  95$&$ 1$&$ 2$&$ 3$&$  0$&$  0.66449 $ \\
		$ 22$&$ 2$&$ 7$&$ 1$&$  0$&$ -0.90298 $&$  59$&$ 0$&$ 7$&$ 6$&$  5$&$  0.39817 $&$  96$&$ 1$&$ 2$&$ 2$&$  1$&$ -0.73483 $ \\
		$ 23$&$ 2$&$ 7$&$ 5$&$  0$&$ -0.29678 $&$  60$&$ 1$&$ 6$&$ 7$&$  0$&$  0.66449 $&$  97$&$ 1$&$ 2$&$ 5$&$  2$&$  0.29678 $ \\
		$ 24$&$ 2$&$ 7$&$ 4$&$  1$&$  0.29678 $&$  61$&$ 1$&$ 6$&$ 6$&$  1$&$  0.45383 $&$  98$&$ 1$&$ 2$&$ 4$&$  3$&$ -0.39817 $ \\
		$ 25$&$ 2$&$ 7$&$ 7$&$  2$&$ -1.48990 $&$  62$&$ 1$&$ 6$&$ 7$&$  4$&$  0.13272 $&$  99$&$ 2$&$ 5$&$ 3$&$  0$&$ -0.13272 $ \\
		$ 26$&$ 2$&$ 7$&$ 6$&$  3$&$  0.64721 $&$  63$&$ 1$&$ 6$&$ 6$&$  5$&$ -0.17807 $&$ 100$&$ 2$&$ 5$&$ 2$&$  1$&$  0.29678 $ \\
		$ 27$&$ 2$&$ 7$&$ 5$&$  4$&$ -0.30186 $&$  64$&$ 4$&$ 7$&$ 7$&$  0$&$ -0.29678 $&$ 101$&$ 2$&$ 5$&$ 5$&$  2$&$ -0.49736 $ \\
		$ 28$&$ 3$&$ 6$&$ 1$&$  0$&$  0.90298 $&$  65$&$ 4$&$ 7$&$ 6$&$  1$&$  0.13272 $&$ 102$&$ 2$&$ 5$&$ 4$&$  3$&$  0.46335 $ \\
		$ 29$&$ 3$&$ 6$&$ 5$&$  0$&$ -0.05936 $&$  66$&$ 4$&$ 7$&$ 7$&$  4$&$ -0.49736 $&$ 103$&$ 3$&$ 4$&$ 3$&$  0$&$  0.17807 $ \\
		$ 30$&$ 3$&$ 6$&$ 4$&$  1$&$  0.05936 $&$  67$&$ 4$&$ 7$&$ 6$&$  5$&$  0.46335 $&$ 104$&$ 3$&$ 4$&$ 2$&$  1$&$ -0.39817 $ \\
		$ 31$&$ 3$&$ 6$&$ 7$&$  2$&$  0.64721 $&$  68$&$ 5$&$ 6$&$ 7$&$  0$&$  0.39817 $&$ 105$&$ 3$&$ 4$&$ 5$&$  2$&$  0.46335 $ \\
		$ 32$&$ 3$&$ 6$&$ 6$&$  3$&$ -0.93736 $&$  69$&$ 5$&$ 6$&$ 6$&$  1$&$ -0.17807 $&$ 106$&$ 3$&$ 4$&$ 4$&$  3$&$ -0.77364 $ \\
		$ 33$&$ 3$&$ 6$&$ 5$&$  4$&$  0.85443 $&$  70$&$ 5$&$ 6$&$ 7$&$  4$&$  0.46335 $&$ 107$&$ 0$&$ 2$&$ 2$&$  0$&$  0.75100 $ \\
		$ 34$&$ 4$&$ 5$&$ 1$&$  0$&$ -0.90298 $&$  71$&$ 5$&$ 6$&$ 6$&$  5$&$ -0.77364 $&$ 108$&$ 2$&$ 4$&$ 4$&$  2$&$ -0.15200 $ \\
		$ 35$&$ 4$&$ 5$&$ 5$&$  0$&$  0.23742 $&$  72$&$ 1$&$ 7$&$ 7$&$  1$&$  0.75100 $&$ 109$&$ 2$&$ 3$&$ 3$&$  2$&$ -0.15200 $ \\
		$ 36$&$ 4$&$ 5$&$ 4$&$  1$&$ -0.23742 $&$  73$&$ 5$&$ 7$&$ 7$&$  5$&$ -0.15200 $&$ -  $&$- $&$ -$&$ -$&$  -$&$       -  $ \\
		\hline
		\bottomrule
	\end{tabular}
\end{table*}

\section{Pivot choices}

Our results in this work are obtained using single-configuration pivots. While a single-configuration pivot is easy to prepare within our hybrid scheme, some states may be absent from the resultant spectral function [e.g., the odd-$J$ states in Fig. \ref{fig:combined_pic}(f) in the main text]. To uncover these missing odd-$J$ states, we can use an alternative single-configuration pivot. More generally, we can prepare a superposition of simple pivots, e.g., by a linear combination of states \cite{lin2022lecture}, to resolve all the eigenstates simultaneously.

\end{document}